\documentclass[conference]{IEEEtran}
\IEEEoverridecommandlockouts
\usepackage{cite}
\usepackage{amssymb,amsfonts}
\usepackage{amsmath}
\usepackage{algorithm}
\usepackage{algpseudocode}
\usepackage{graphicx}
\usepackage{textcomp}
\usepackage{xcolor}
\usepackage{fancyhdr}
\usepackage[normalem]{ulem}
\usepackage{algpseudocode}
\usepackage{subfigure}
\usepackage{cancel}
\def\BibTeX{{\rm B\kern-.05em{\sc i\kern-.025em b}\kern-.08em
    T\kern-.1667em\lower.7ex\hbox{E}\kern-.125emX}}

\newboolean{showcomments}
\setboolean{showcomments}{true}
\ifthenelse{\boolean{showcomments}}
{
}

\newcommand*{\affmark}[1][*]{\textsuperscript{#1}}

\begin{document}

\title{Transformer-Based Wireless Traffic Prediction and Network Optimization in O-RAN}

\author{\IEEEauthorblockN{Md~Arafat~Habib\affmark[1], Pedro~Enrique~Iturria-Rivera\affmark[1], Yigit Ozcan\affmark[2], Medhat Elsayed\affmark[2], \\Majid Bavand\affmark[2], Raimundus Gaigalas\affmark[2] and  Melike Erol-Kantarci\affmark[1], \IEEEmembership{Senior Member,~IEEE}}
\IEEEauthorblockA{\affmark[1]\textit{School of Electrical Engineering and Computer Science, University of Ottawa, Ottawa, Canada}}  \affmark[2]\textit{Ericsson Inc., Ottawa, Canada}\\
Emails:\{mhabi050, pitur008, melike.erolkantarci\}@uottawa.ca, \\\{yigit.ozcan, medhat.elsayed, majid.bavand, raimundas.gaigalas \}@ericsson.com \vspace{-1em}}

\maketitle

\thispagestyle{fancy}   
\fancyhead{}                
\lhead{Accepted by the WS03 ICC 2024 Workshop \copyright2023 IEEE}
\cfoot{}
\renewcommand{\headrulewidth}{0pt} 

\begin{abstract}
This paper introduces an innovative method for predicting wireless network traffic in concise temporal intervals for Open Radio Access Networks (O-RAN) using a transformer architecture, which is the machine learning model behind generative AI tools. Depending on the anticipated traffic, the system either launches a reinforcement learning-based traffic steering xApp or a cell sleeping rApp to enhance performance metrics like throughput or energy efficiency. Our simulation results demonstrate that the proposed traffic prediction-based network optimization mechanism matches the performance of standalone RAN applications (rApps/ xApps) that are always on during the whole simulation time while offering on-demand activation. This feature is particularly advantageous during instances of abrupt fluctuations in traffic volume. Rather than persistently operating specific applications irrespective of the actual incoming traffic conditions, the proposed prediction-based method increases the average energy efficiency by $39.7\%$ compared to the `Always on Traffic Steering xApp' and achieves $10.1\%$ increase in throughput compared to the `Always on Cell Sleeping rApp'. The simulation has been conducted over $24$ hours, emulating a whole day traffic pattern for a dense urban area. 
\end{abstract}


\section{Introduction}
\label{s1} 

Estimating future traffic volume accurately in a time-sensitive manner is extremely advantageous for contemporary 5G network infrastructures. Based on the predicted traffic pattern, one can design and initiate various network functions for performance optimization for different traffic volumes. For example, whenever the traffic volume is predicted to be high, one can perform traffic steering (designed as an xApp in Open Radio Access Network (O-RAN) terminology) to maintain optimal throughput for enforcing stringent Quality-of-Service (QoS) requirements. However, predicting 5G traffic poses a substantial difficulty due to its highly variable nature which adds complexity to the forecasting process. The difficulty is further exacerbated by the challenge of making real-time predictions that depend on long-term data trends. Additionally, several performance-enhancing xApps, such as power allocation, scheduling, and traffic steering, necessitate a more detailed prediction of traffic, ideally at the sub-second level, to effectively implement these applications in a proactive manner based on such forecasts.

5G networks feature short Transmission Time Intervals (TTIs), increasing data processing needs and sensitivity to traffic fluctuations. These networks cater to various services like Enhanced Mobile Broadband (eMBB), Ultra-Reliable Low-Latency Communications (URLLC), and Massive Machine-type Communications (mMTC), each with unique, rapidly changing traffic patterns \cite{19,20}. Accurately predicting these patterns is challenging due to the complex temporal correlations in traffic data at the TTI level, encompassing both short and long-term dependencies. Therefore, a sophisticated traffic prediction system is essential for effectively managing these dynamics in 5G networks. 

Advanced machine learning-based time series forecasting methods have been used in the literature to predict network traffic in the temporal domain. Most of the works in the literature are limited to Recurrent Neural Network (RNN)/ Long Short-Term Memory (LSTM) type methods. For instance, Trinh et al. propose an LSTM-based mobile traffic prediction using raw data \cite{1}. However, LSTMs are known to struggle with capturing long-term dependencies in sequences. As the sequence length increases, the ability of LSTMs to retain and propagate information over long distances in the sequence diminishes. It's also worth noting that while the literature contains studies on wireless traffic prediction, the integration of these techniques within the O-RAN framework, particularly for prediction-led network optimization in O-RAN, is an area that remains unexplored.  
  
To this end, we propose to use a transformer-based method named Autoformer \cite{2} for wireless traffic prediction in concise temporal intervals. Based on the predicted traffic volume, we fix thresholds for high and low traffic loads leading to xApp or rApp orchestration (initialization/ termination) for performance optimization of multiple conflicting Key Performance Indicators (KPIs). Different from the previous works, the proposed scheme provides a time window to initiate network functions in advance to tackle upcoming heavy-weight traffic data more efficiently. Instead of keeping a certain network-optimizing functionality always turned on, we trigger it on-demand based on the predicted traffic load. This saves processing cycles for network optimization.

The results from our simulations indicate that the network optimization mechanism we have developed performs on par with standalone xApps/rApps that get continuously executed. This mechanism is particularly effective in responding to sudden changes in traffic volume, as it can be activated on demand. Unlike traditional methods that constantly run certain applications (like the cell sleeping rApp or traffic steering xApp) regardless of the real-time traffic situation, our predictive approach enhances energy efficiency by an average of $39.7\%$ compared to the `Always on Traffic Steering xApp'. It also improves throughput by $10.1\%$ over the `Always on Cell Sleeping rApp'. These simulations were conducted over 24 hours to replicate daily traffic patterns in a densely populated urban area. 

The structure of this paper is as follows: Section \ref{s2} provides an overview of the literature related to this topic. This is followed by Section \ref{s3}, which details the system model extensively. Section \ref{s4} is dedicated to explaining our transformer-based network optimization approach in O-RAN. The performance evaluation and comparative analysis of our method against established baselines are detailed in Section \ref{s5}. Finally, Section \ref{s6} concludes the paper.

\section{Related work}
\label{s2}

Some traditional time series forecasting methods like Holt-Winters and seasonal autoregressive integrated moving averages have been used for wireless traffic prediction in mobile networks \cite{3,4}. However, as the traffic volume has surged highly after the emergence of the 5G era, advanced machine learning models like convolutional neural networks, RNNs, and LSTMs have been used to explore the long-term temporal dependencies of traffic.  

An RNN utilizing LSTM units for extracting temporal dependencies is utilized in \cite{5}, where an attention mechanism has also been introduced to enhance prediction accuracy by focusing on significant features. A deep learning framework designed to improve cellular traffic prediction accuracy is proposed in \cite{6}, having a spatial-temporal attention mechanism integrated into an LSTM model. Another LSTM-based traffic prediction framework is presented in \cite{7} to tackle diverse and dynamic traffic demands in 5G O-RAN. 

The authors of  \cite{8}  proposed to address the increasing complexity and data volume in 5G networks by proposing a novel deep learning method that utilizes a lightweight hybrid attention network. It combines efficient hybrid attention mechanisms with depthwise separable convolutions to improve feature extraction while maintaining lower computational costs.  

Compared with existing literature, the main contribution of this work is that we propose a transformer-based wireless traffic prediction in the temporal domain. The prediction is done in finer time frames and based on the predicted traffic volume, network optimization in O-RAN is performed via traffic steering and cell sleeping Apps. To the best of our knowledge, this is the first attempt in the literature to predict wireless traffic in such granular time frames using Transformer.

\section{System Model}
\label{s3}
\subsection{System Model}
\begin{figure}[!t]
\centerline{\includegraphics[width=0.7\linewidth]{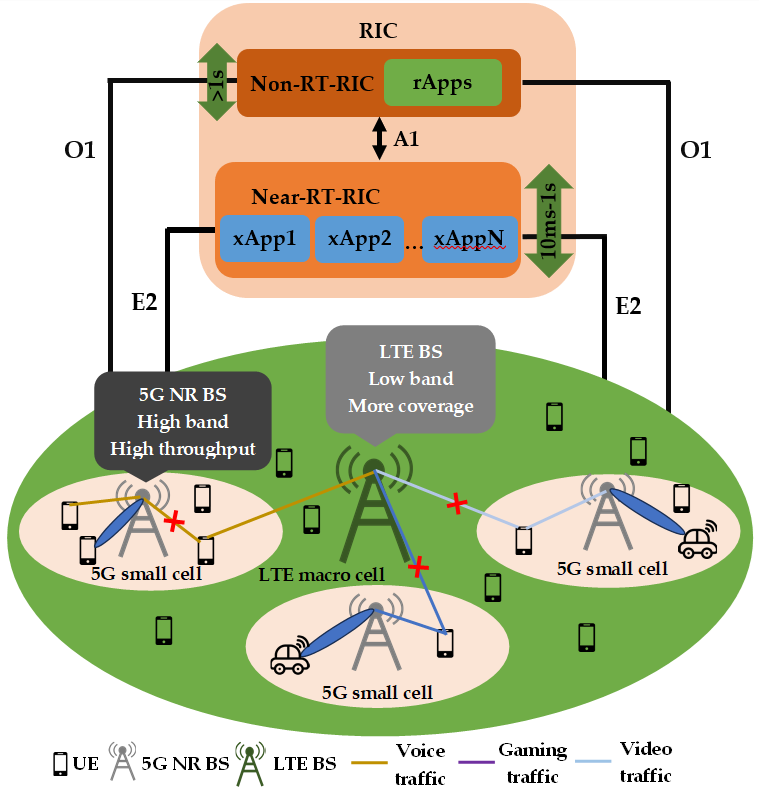}}
\caption{O-RAN based network stack with macro cell and small cells based deployment.}
\label{fig1}
\vspace{-1.2em}
\end{figure}

We consider an O-RAN-based downlink orthogonal frequency division multiplexing cellular system having $B$ BSs serving $U$ users simultaneously, and multiple small cells in the system are within the range of a macro cell. There are $\kappa$ classes of traffic in the system and users are connected with multiple RATs via dual connectivity. There are $Q$ classes of RATs ($q_1,q_2,...,q_n$), where $q$ represents a certain access technology (LTE, 5G, etc.). The wireless system model considered in this work is presented in Fig. \ref{fig1}. The RAN Intelligent Controllers (RICs) depicted in the figure, namely Non-Real-Time (non-RT)-RIC and Near-Real-Time (Near-RT)-RIC, are capable of hosting rApps and xApps. These applications, which focus on control and optimization, operate across various time scales according to O-RAN architecture.

In our proposed system, we incorporate an advanced machine learning-based aggregate traffic predictor that resides in the near-RT-RIC as an xApp. It can predict upcoming traffic based on past traffic patterns. The traffic is forecast for successive time intervals, and two thresholds are established based on the training data for the model. The first threshold, $Th_p$, is determined from past measurements of high traffic volumes. When the predicted traffic exceeds $Th_p$, applications like traffic steering, power allocation, and others that can increase/ maintain high throughput can be activated. The second threshold, $Th_t$, is derived from low traffic volumes. If the predicted traffic falls below this threshold, energy-saving rApp like cell sleeping can be initiated to optimize energy usage due to the expected lower traffic levels. A feedback loop is implemented for dynamically adjusting $Th_p$ and $Th_t$. It provides performance feedback through the O1 interface. The feedback, resulting from shifts in data, may necessitate the reevaluation of thresholds. If deemed necessary, the transformer model can be retrained with the updated data.    

\subsection{Problem formulation}

Let's assume $A$ represents the set of all xApps and rApps, $B$ is a subset of $A$ ($B\subseteq A$) containing at least one element (an xApp or rApp) capable of enhancing network performance based on the predicted traffic volume. Let $C$ be the set of potential KPIs that an xApp or an rApp might improve. Under these assumptions, the problem of optimizing network performance can be formulated as follows:

\begin{equation} \label{eq5}
    \begin{split}
     max \sum_{c \in C}\{ P_{t+1(c)}-P_{t(c)}\}, \quad \quad\quad \\
     s.t. \quad 
     \forall(A)\exists(B): W(Q) = 1, \quad \quad  \\
     S + V \leq 1, \quad \quad \quad \quad \quad
    \end{split}
\end{equation}
where $P_{t(c)}$ is the value of a certain performance metric, $c$ (e.g. throughput, energy efficiency) at time $t$,  $P_{t+1(c)}$ is the value of the same performance metric at time $t+1$. The objective of the proposed system is to predict traffic, $T_p$ at time $t$, and based on $T_p$, initiate an xApp or rApp to maximize a performance metric $c$ at $t+1$. We want to optimize $c$, subject to the condition $S + V \leq 1$, which ensures that conflicting xApps or rApps cannot be active simultaneously. This prevents conflicting actions from being taken by different xApps and rApps that are designed for opposite traffic conditions. In the condition, $S$ represents traffic steering xApp, where $S=1$ if the xApp is active and $S=0$ otherwise. Also, $V$ represents the cell sleeping rApp, where $V=1$ if it is active and $V=0$ otherwise. Furthermore, $W(Q)$ is a proposition stating ‘An xApp can improve a performance metric’, which is either ‘0’ or ‘1’.    

\section{Proposed Transformer-based Traffic Prediction and Performance Optimization Scheme}
\label{s4}

In this section, first, we provide the details of the data collection mechanism to be used for traffic prediction. Next, the traffic predictor xApp and associated RIC applications are discussed in brief. Finally, we present the proposed Autoformer-based traffic prediction mechanism and network optimization processes in detail.

\subsection{Data Aggregation and Pre-processing}

Based on the presented multi-RAT environment with dynamic traffic classes (Fig. \ref{fig1}), we collected simulation data for training our traffic predictor. We have followed the method presented in \cite{1} to collect data for multiple TTIs. The data collection model used in the system is as follows: 
\begin{equation} \label{eq1}
    \bf{D}=\{D_{y_1}, D_{y_2},...,D_{y_z}\},
\end{equation}
where $D_{y_{z}}$ is a set of measurements for a monitored cell $z$. We store this information in $s_r(t)$, where $t$ corresponds to a TTI, which is 1ms ($\mu = 0$) long in our case. We calculate the aggregate cell traffic measurements for a given time frame, $T$ (multiple TTIs).
\begin{equation} \label{eq2}
     \bf{S}(T)=\sum_{t\in T} s_r(t), 
\end{equation}
where $\bf{S(T)}$ is the sum of the traffic generated by all the UEs connected during the time frame $T$. For the one-step prediction, we use a many-to-one architecture \cite{1}. This means that the network observes the mobile traffic for a fixed number of timeslots until $T$ and, then tries to predict the traffic in the next time slot $T+1$.

We have used the Savitzky-Golay smoothing filter \cite{17} technique to smoothen our data so that higher moment features in the data, such as the width and height of peaks, and overall trend are preserved. It effectively reduces noise while maintaining the structure of the signal, which is important for identifying the underlying trends in the data without being misled by random fluctuations. 

\subsection{Traffic Prediction xApp}

To perform traffic prediction in TTI level time granularity, we have used Autoformer \cite{2}. This model is designed to address challenges in forecasting intricate temporal patterns over extended periods, which LSTM and traditional Transformer-based models struggle with due to computational limitations and inefficiencies in capturing long-range dependencies. Data collected using the method described in Section \ref{s4}A are passed to the Autoformer as inputs. 

The core innovation of the Autoformer lies in its unique decomposition architecture and the Auto-Correlation mechanism. This design enables the model to effectively decompose the huge blocks of temporal domain traffic data into more manageable components and leverage the inherent periodicity in the data for more efficient and accurate forecasting.  

The decomposition block within Autoformer architecture is designed to progressively isolate the long-term stable trend from the evolving hidden variables in its predictions. Specifically, it employs a modified moving average technique that effectively dampens periodic variations, thereby accentuating the underlying long-term trends. For traffic data input, $D$, the process can be written as: 
\begin{equation} \label{eq3}
    \begin{split}
        D_t=AvgPool(Padding(D)) \\ 
        D_s=D-D_t, \quad \quad \quad
    \end{split}
\end{equation}
where $D_s$ and $D_t$ denote the seasonal and the extracted trend cycle parts respectively. The $AvgPool(.)$ is adopted for the moving average with the padding operation to keep the series length unchanged. 

\subsection{Traffic Steering and Cell Sleeping Apps}

The effectiveness of the proposed method is demonstrated through the use of a traffic steering xApp and a cell sleeping rApp in this paper. The traffic steering xApp is designed to simultaneously maintain the QoS for different types of traffic by implementing a steering mechanism using a Deep Q-Network (DQN). To ensure effective performance, the xApp's state and reward functions are tailored with an emphasis on two Key Performance Indicators (KPIs): network latency and overall system throughput. It directs traffic toward specific Base Stations (BSs) considering factors such as load at each BS, link quality, and the nature of the traffic. For more in-depth information on this xApp, refer to \cite{12}.

The cell sleeping rApp aims to reduce network power consumption by deactivating idle or less busy BSs based on traffic load and queue length. It models BS power consumption with a fixed rate and an additional variable rate based on transmission power. In sleep mode, a BS consumes less power. The xApp's goal is to maximize energy efficiency without overloading active BSs. It does this by optimizing a formula that considers each BS's total throughput per power unit, deducting penalties for any overloaded BS to avoid strain on active ones. Technical details of this DQN-based xApp can be found in one of our previous works \cite{16}.

\subsection{Network Optimization Using Autoformer and RL-based RIC applications}

The first step of the proposed method is to collect cell traffic data using the data aggregation method provided in Section \ref{s4}A. Next, data is pre-processed before providing as inputs to Autoformer to forecast traffic for successive time intervals. Two thresholds, $Th_t$ and $Th_p$ are defined based on the past traffic volume. The traffic volume can fluctuate significantly during the day depending on user activity. We define $Th_p$ as the threshold of traffic volume above which we consider the traffic to be very high, necessitating performance optimization to manage optimal throughput. On the other hand, $Th_t$ is the threshold below which traffic volume is deemed low enough to activate an energy-saving cell sleeping application. By reviewing past performance data for various traffic volumes, we aim to determine optimal threshold values. However, these thresholds can be set by using a semi-supervised learning approach for real deployments. If we set a specific value as $Th_p$, and the predicted incoming traffic volume exceeds this value, the traffic steering application is triggered. Similarly, if the predicted traffic falls below $Th_t$, the cell sleeping application is initiated to save energy.

\begin{figure}[!t]
\centerline{\includegraphics[width=0.7\linewidth]{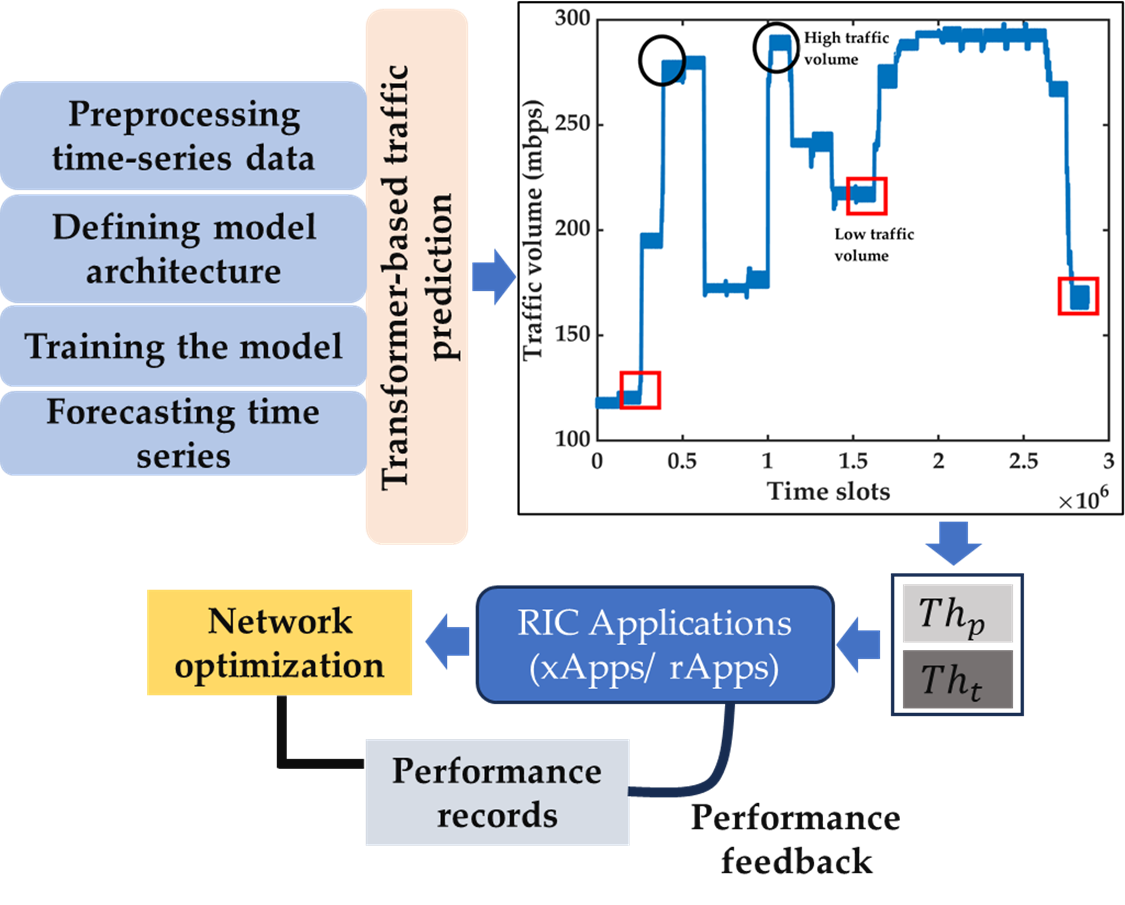}}
\caption{Traffic prediction and network optimization based on the proposed method. }
\vspace{-1.2em}
\label{fig4}
\end{figure}

As presented in Fig. \ref{fig4}, based on the threshold, an xApp or an rApp can be initiated/ terminated. The whole process of traffic prediction-based network optimization using transformer and Reinforcement Learning (RL) can be summarized as follows: 
\begin{itemize}
\item \textbf{Step 1:} Simulation data collection and pre-processing of the data is performed using the Savitzky-Golay filter.   
\item \textbf{Step 2:} In the next step, as presented in \ref{fig4}, we define our transformer model, which is Autoformer in our case. 
\item \textbf{Step 3:} We train the Autoformer model with the past time series data and decide $Th_p$ and $Th_t$.  
\item \textbf{Step 4:} In this step, based on the anticipated traffic, a decision is made whether to initiate an RL-based RIC application (xApp or rApp) in advance to comply with the time duration left to tackle the upcoming traffic volume. 
\item \textbf{Step 5:} The performance based on the crucial KPIs is recorded and fed back to the system for the adjustment of the thresholds and input data to the Autoformer for precise and robust prediction. 
\end{itemize}

\subsection{Baseline Algorithms}

To show the robustness and the importance of the proposed method, we have compared the performance with two different baselines. The first one is `Always on Traffic Steering xApp', where we run the traffic steering xApp for the whole simulation time of 24 hours. We compare the performance of the proposed prediction-led method with this standalone scenario to show that the on-demand activation/ termination of traffic steering xApp leads to better energy efficiency in situations when we have varying traffic volume in the temporal dimension. Similarly, the second baseline is `Always on Cell Sleeping rApp'. 

Furthermore, to demonstrate the fact that the proposed Autformer-based prediction method works better than the existing state-of-the-art time series prediction algorithms, we have done a performance comparison with two more baselines. One is LSTM-based wireless traffic prediction \cite{1} using the same simulation data that have been used for the proposed method. The second baseline is inspired by Generative Pre-Trained Transformer version 2 (GPT2), where a Decoder-only Transformer (DOT) has been used \cite{13}.

\section{Performance Evaluation}
\label{s5}
\subsection{Simulation setup}
We have used the MATLAB 5G Toolbox with O-RAN 7.2 split for simulation. Training is performed offline to establish the model, followed by online testing. In the simulation setup, there is one macro-cell served by an eNB and four small cells served by gNBs, accommodating 60 users. The simulation involves three kinds of traffic: Video, Gaming, and Voice. The QoS demands for each traffic type adhere to the standards outlined in 3GPP and additional specifications cited in \cite{14}. Voice traffic is characterized by a throughput requirement of 0.1 Mbps and a delay of 100 ms. Video traffic, in contrast, has parameters set at 10 Mbps for throughput, and an 80 ms delay. Gaming traffic has a throughput requirement of 5 Mbps and a delay of 40 ms. The simulation operates in a 5G NSA mode \cite{18}, integrating different RATs (LTE and 5G). Table \ref{tab1} summarizes the simulation settings along with the Autoformer and RL parameters. 

\begin{table}[htp]
    \centering
    \caption{Simulation settings for multi-RAT Implementation}
    \begin{tabular}{|l|l|}
         \hline
         \textbf{\underline{5G NR}} & \\
         Bandwidth & $20$ MHz \cite{18}\\
         Carrier frequency &  $3.5$ GHz \cite{18}\\
         Subcarrier spacing & $15$ KHz \\
         Max transmission power & $43$ dBm \cite{15}\\
         Channel model & 3GPP urban micro\\
         \hline
         \textbf{\underline{LTE}} & \\
         Bandwidth & $10$ MHz \cite{18}\\
         Carrier frequency & $800$ MHz \cite{18}\\
         Subcarrier spacing & $15$ KHz \\
         Max transmission power & $38$ dBm \cite{15}\\
         Channel model & 3GPP urban macro\\
         \hline
         \textbf{\underline{Video, Gaming, and voice traffic}} & \\
         Inter-arrival time between packets & 12.5, 40, and 0.1 ms \cite{14}\\ 
         Distribution & Pareto, Uniform, and Poisson \cite{14}\\
         Packet size & $250, 120,$ and $ 30$ bytes \cite{14}\\
         \hline
         \textbf{\underline{RL parameters}} & \\
         Batch size, Initial exploring steps & $32$, $3000$ \\
         Target network frequency updates & $1000$ \\
         Learning rate ($\alpha$), discount factor $\gamma$ & $0.5$, $0.9$ \\
         \hline
         \textbf{\underline{Autoformer}} & \\
         Learning rate $(\alpha)$ & $10^{-4}$ \cite{2} \\
         Batch size & $32$ \cite{2}\\
         Model architecture & 2 encoder layers, 1 decoder layer\\
         Auto-correlation hyperparameter ($\rho$) & 2\\
         \hline
    \end{tabular}
    \label{tab1}
\end{table}

We have performed a two-day long simulation to collect data as input to the Autoformer (24 hours) and predict (next 24 hours). Based on the traffic data in our simulation, we have observed that the traffic volume can reach as high as $298$ Mbps for $60$ users, $5$ cells and go as low as $116$ Mbps during off-peak hours. We observe the past performance graph for different traffic volumes (e.g. $120, 140, 190, 300,...$) and decide what should be an optimal threshold $Th_p$ and $Th_t$. $Th_p$ is set to be $220$ Mbps, which means whenever the predicted value of the incoming traffic is anticipated to go beyond this value, traffic steering xApp is initiated. Similarly, if the predicted traffic goes below $Th_t$, we can initiate cell sleeping rApp.

\subsection{Simulation results}
\begin{figure}[!t]
\centerline{\includegraphics[width=0.65\linewidth]{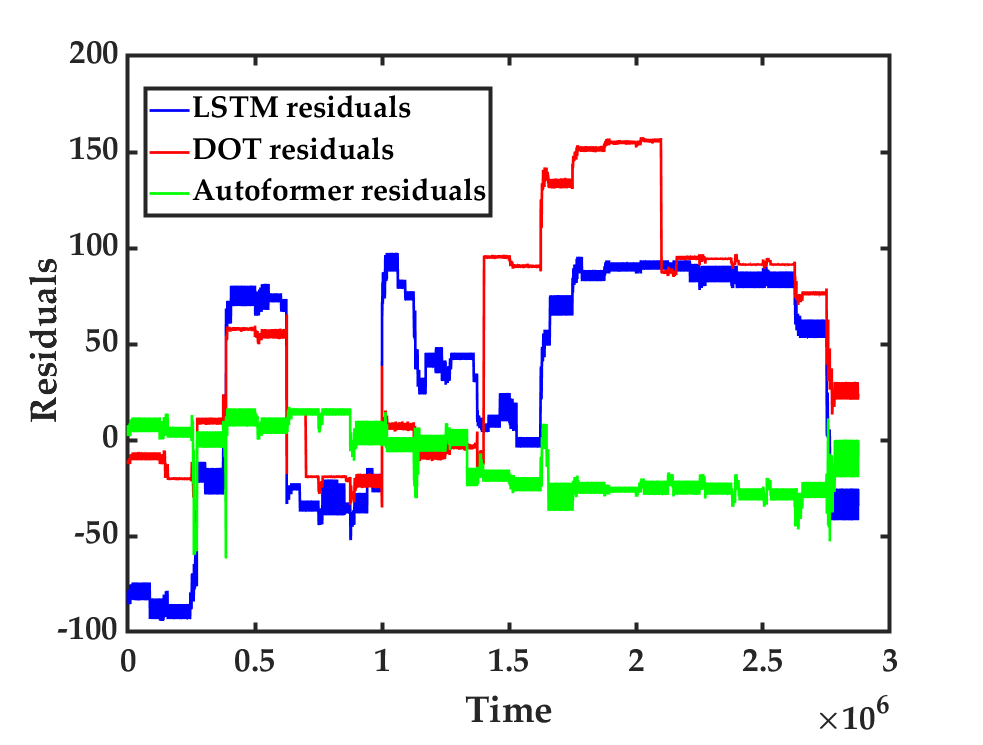}}
\caption{Residual plots of the proposed method and the baselines.}
\label{fig5}
\vspace{-1.2em}
\end{figure}

First, we present a performance comparison among the prediction methods: the proposed Autoformer-based wireless traffic prediction against the two baselines: LSTM, and GPT2's DOT. We provide the comparison based on time series residual plots. Residuals are obtained by subtracting the predicted values from the actual observed values in the time series. The formula to calculate residuals ($e_t$) for a time series is: 
\begin{equation}
    e_t = y_t - \hat{y}_t,
\end{equation}
where $e_t$ is the residual at time $t$, $y_t$ is the actual value observed at time $t$, and $\hat{y}_t$ is the predicted value at time $t$, given by the model. Residuals are plotted against time and a consistent pattern (e.g., increasing or decreasing residuals) suggests that the model has failed to capture certain temporal structures in the data. As we can see from Fig. \ref{fig5}, residuals of the proposed Autoformer-based wireless traffic prediction are scattered around zero whereas residuals of the baselines fluctuate highly.

\begin{figure}[!t]
\centerline{\includegraphics[width=0.65\linewidth]{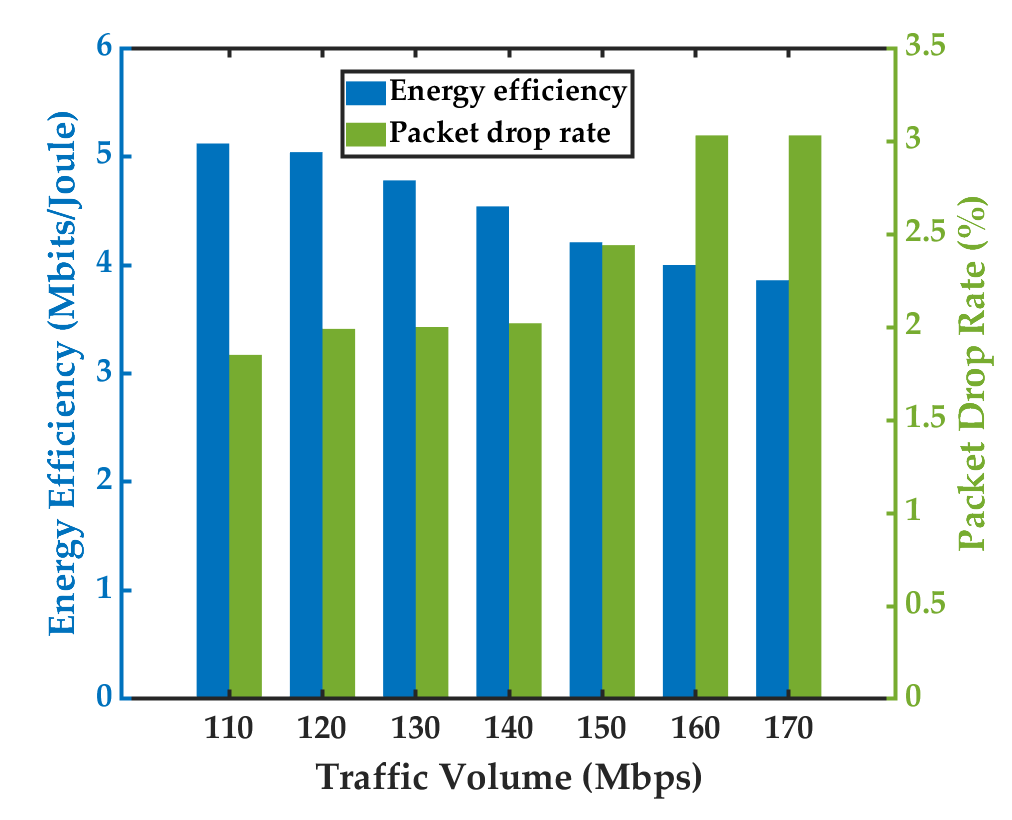}}
\caption{Threshold value selection for initiating cell sleeping rApp.} 
\label{fig6}
\vspace{-1.2em}
\end{figure}

Fig. \ref{fig6} presents a bar plot that provides with visual representation of concurrent energy efficiency and packet drop rate. Based on the data presented in Fig. \ref{fig6}, at around $140$ Mbps traffic volume, the energy efficiency shows a more significant drop (from $4.54$ to $4.21$ Mbits/Joule) without a dramatic increase in packet drop rate. Therefore, the threshold ($Th_t$) is set to be $140$ Mbps. If the anticipated traffic volume goes below this threshold, cell sleeping rApp is initiated. The threshold ($Th_p$) for initiating traffic steering xApp is decided similarly for higher traffic volume using performance graphs.   

\begin{figure}[!t]
\centerline{\includegraphics[width=0.65\linewidth]{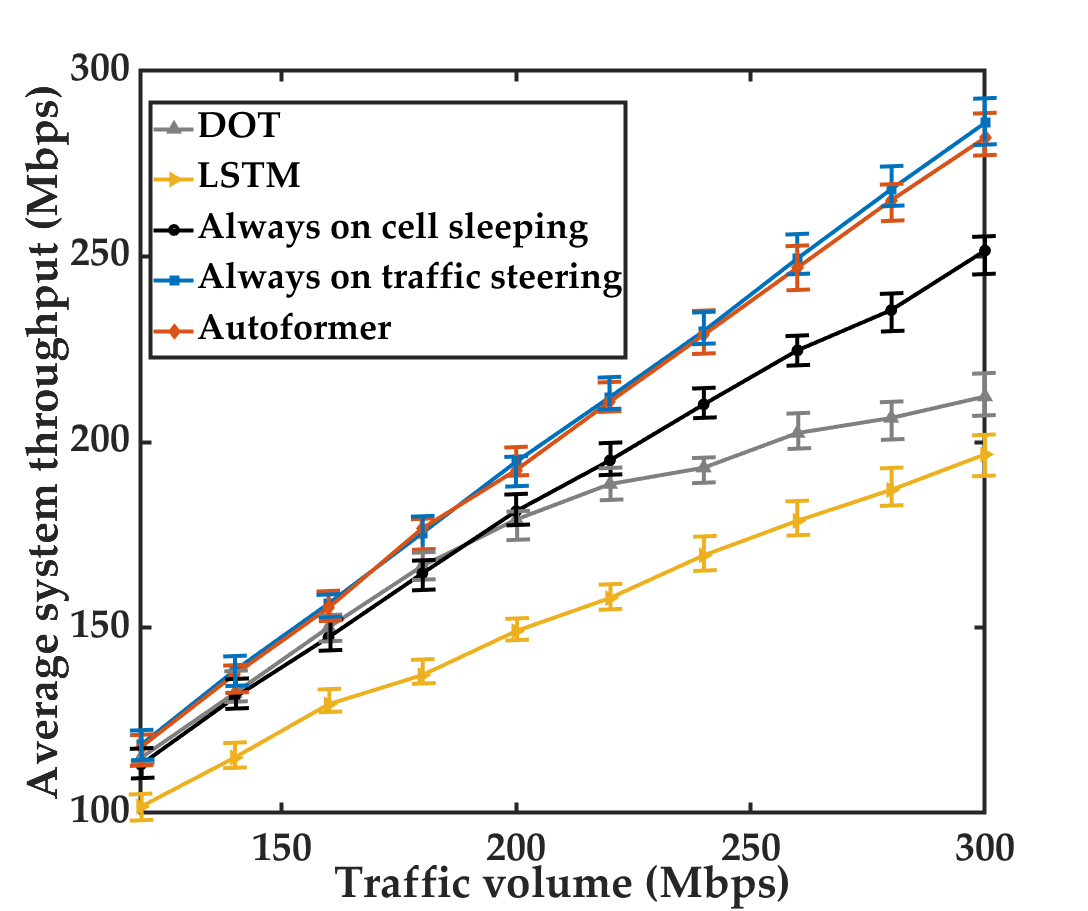}}
\caption{Performance comparison in terms of throughput.}
\label{fig7}
\vspace{-1.2em}
\end{figure}

\begin{figure}[!t]
\centerline{\includegraphics[width=0.65\linewidth]{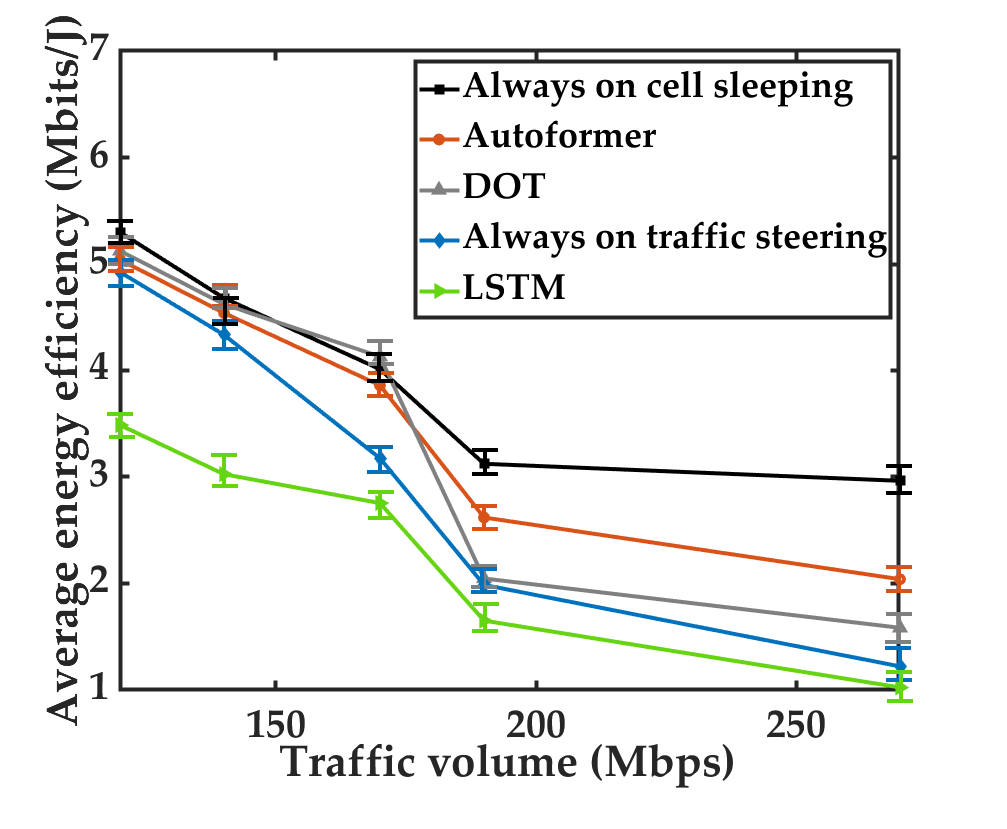}}
\caption{Performance comparison in terms of energy efficiency.}
\label{fig8}
\vspace{-1.2em}
\end{figure}

Next, we show the superior performance achievement of the proposed wireless traffic prediction and network optimization method. Autoformer-based predictor gives $23.35\%$ and $28.44\%$ increase in throughput (Fig. \ref{fig7}) compared to LSTM and DOT respectively. It can be seen from Fig. \ref{fig7} that it achieves similar performance in comparison to the `Always on Traffic Steering' scenario. Keeping the traffic steering xApp always turned on no matter what the traffic volume is  (low/ high) comes with a significant energy cost. This is because traffic steering xApp would require more cells to be active even when there is not enough traffic load in the off-peak hours. This is where the proposed on-demand Autoformer-based wireless traffic prediction along with RL-based network optimization plays a crucial role. Since the method is on demand, it is way more energy efficient than running the traffic steering xApp always. Fig. \ref{fig8} presents us with a vivid comparison between the proposed method and the baseline algorithms in terms of energy efficiency. It is visible that the `Always on Traffic Steering' scheme in this case performs poorly compared to the proposed method. Furthermore, the proposed method outsmarts the baseline algorithms (LSTM and DOT) in terms of energy efficiency by achieving a $22.54\%$ and $49.8\%$ increase. Note that in both Fig.\ref{fig7} and \ref{fig8}, traffic volume in the X-axis is average but not always exact.  

To summarize, the proposed method surpasses two key benchmark algorithms in the literature, demonstrating superior performance at both prediction and network levels. LSTM fails in handling long-term dependencies in sequences as its efficiency decreases with longer sequences. Transformer, conversely, is adept at managing long-range dependencies due to its self-attention mechanism, which allows it to focus on any part of the input sequence. Our method employs Autoformer, a transformer-based approach, which uniquely separates time series data into trend-cyclical and seasonal parts for better long-term forecasting. This decomposition enables precise extraction of long-term trends resulting in better performance compared to GPT-2's DOT. Lastly, due to its on-demand activation policy, it performs significantly better than stand-alone xApp scenarios where incoming traffic fluctuates in the long run. 

\section{Conclusions}
\label{s6}

In this work, we have developed an innovative method for the prediction of wireless traffic in short time intervals within O-RAN, utilizing a transformer-based architecture. Depending on the forecast traffic, our system dynamically orchestrates and deploys either an RL-driven traffic steering xApp or a cell sleeping rApp.  Through rigorous simulation, our method has proven to match the efficiency of continuously running standalone xApps or rApps, while also providing the flexibility of on-demand activation. Unlike the constant operation of specific RIC applications, our predictive approach has been shown to enhance average energy efficiency by 39.7\% and average system throughput by 10.1\% compared to ``Always-on  Traffic Steering and Cell Sleeping Apps''.

\section*{Acknowledgement}
This work has been supported by MITACS and Ericsson Canada, NSERC Canada Research Chairs, and NSERC Collaborative Research program.

\bibliographystyle{IEEEtran}
\bibliography{ref.bib}
\end{document}